\documentclass[conference]{IEEEtran}
\IEEEoverridecommandlockouts
\usepackage{cite}
\usepackage{amsmath,amssymb,amsfonts}
\usepackage{algorithmic}
\usepackage{graphicx}
\usepackage{textcomp}
\usepackage{xcolor}

\usepackage{tikz}

\usepackage{filecontents}
\usepackage[skip=0pt,belowskip=-17pt]{caption}
\usepackage[skip=0pt,belowskip=2pt]{subcaption}
\usepackage[linesnumbered,ruled,vlined,noend]{algorithm2e}
\usepackage[inline]{enumitem}
\usepackage{balance}
\usepackage{amsmath}
\usepackage{amssymb}
\usepackage{titlesec}

\usepackage[all=normal, tracking=tight]{savetrees}

\begin{document}

\title{MSF-Model: Queuing-Based Analysis and Prediction of Metastable Failures in Replicated Storage Systems}

\author{\IEEEauthorblockN{Farzad Habibi}
\IEEEauthorblockA{
\textit{UC Irvine} \\
habibif@uci.edu}
\and
\IEEEauthorblockN{Tania Lorido-Botran}
\IEEEauthorblockA{\textit{Roblox} \\
tbotran@roblox.com}
\and
\IEEEauthorblockN{Ahmad Showail}
\IEEEauthorblockA{\textit{Taibah University}\\
\textit{University of Prince Mugrin} \\
ashowail@taibahu.edu.sa}
\and 
\IEEEauthorblockN{Daniel C. Sturman}
\IEEEauthorblockA{\textit{Roblox} \\
sturman@roblox.com}
\and
\IEEEauthorblockN{Faisal Nawab}
\IEEEauthorblockA{\textit{UC Irvine} \\
nawabf@uci.edu}
}

\IEEEaftertitletext{\vspace{-3\baselineskip}}
\microtypesetup{deactivate}
\maketitle
\microtypesetup{reactivate}

\newcommand{\ourModelName}{MSF-Model}
\newcommand{\ourModel}{\emph{\ourModelName}}
\newcommand{\ourModelFull}{MSF-Model: Queuing-Based Analysis and Prediction of Metastable Failures in Replicated Storage Systems}

\let\oldnl\nl
\newcommand{\nonl}{\renewcommand{\nl}{\let\nl\oldnl}}

\newcommand{\cat}[1]{\smallskip\noindent\textbf{#1.}}
\newcommand{\catnum}[2]{\smallskip\noindent\textbf{#1. #2:}}

%

\begin{abstract}
Metastable failure is a recent abstraction of a pattern of failures that occurs frequently in real-world distributed storage systems. 
%
%
In this paper, we propose a formal analysis and modeling of metastable failures in replicated storage systems. We focus on a foundational problem in distributed systems---the problem of consensus---to have an impact on a large class of systems. 
%
Our main contribution is the development of a queuing-based analytical model, \ourModel, that can be used to characterize and predict metastable failures. \ourModel\ integrates novel modeling concepts that allow modeling metastable failures, which was intractable to model prior to our work. We also perform real experiments to reproduce and validate our model. Our real experiments show that \ourModel\ predicts metastable failures with high accuracy by comparing the real experiment with the predictions from the queuing-based model. 
\end{abstract}

\IEEEpeerreviewmaketitle

\begin{figure*}[h]
  \centering
  
  \begin{minipage}[b]{ 0.31\textwidth}
    \centering
    \includegraphics[width=\linewidth]{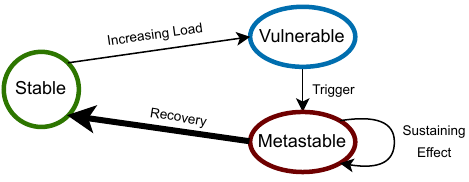}
    \caption{Life cycle of a metastable failure}
    \label{fig:ms-failure-lifecycle}
  \end{minipage}
  \hfill
  \begin{minipage}[b]{0.37\textwidth}
    \centering
    \includegraphics[width=\linewidth]{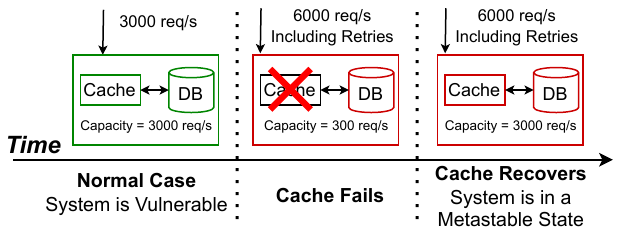}
    \caption{Timeline of a metastable failure scenario}
    \label{fig:ms-timeline}
  \end{minipage}
  \hfill
  \begin{minipage}[b]{0.29\textwidth}
    \centering
    \centering
    \includegraphics[width=\linewidth]{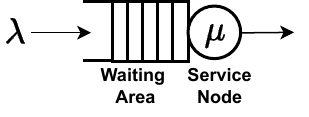}
    \caption{An $M/M/1/\infty$ queuing node}
    \label{fig:mm1}
  \end{minipage}

\end{figure*}

\section{Introduction}


Previous works studied the reliability of distributed storage systems under conditions of hardware failures, software bugs, network outages, and human errors~\cite{schlichting1983fail, lamport2019parttimeParliment, ongaro2014raft,lamport2019byzantine, castro1999practicalByzantine}. 
The rise of cloud services has given rise to new failure types, including security failures~\cite{li2013secFailure1, somorovsky2011secFailure2}, stragglers~\cite{cipar2013straggler1, dean2008straggler2}, configuration failures~\cite{oppenheimer2003confiFailure1, yuan2014configFailure2}, fail-slow failures~\cite{fastFailSlowLu, arpaci2001failslow1, gunawi2016whyCloudStop}, and most recently metastable failures introduced by Bronson et al.~\cite{bronson2021metastable} and explored further by Huang et al.~\cite{huang2022metastable}.

\cat{Metastable Failures}
A metastable failure~\cite{bronson2021metastable, huang2022metastable} describes a state of the system that---although functioning---has extremely low performance due to \emph{sustained artificial overload}. We emphasize the word \emph{artificial} as the overload is not solely due to incoming traffic (though it might be \emph{triggered} initially by it); rather, the overload is caused by an artifact of the system. 
An example of a metastable failure is a \emph{never-ending} feedback loop of retrial requests (called \emph{retry storm}) that is triggered initially by a \emph{temporary} sudden sharp increase in incoming traffic. The temporary incoming traffic overload stresses the system leading to delays in processing requests and leads to triggering retrial requests. What makes it a metastable failure is that the retrial requests themselves cause a further increase in the load on the system; this leads to a feedback loop of retrial requests increasing the load on the system, leading to more retrial requests, and so on. At this stage, even if the incoming traffic is reduced to normal levels, the overload from retrial requests is continuous and high enough to (artificially) sustain the overload. 

Metastable failures have been recurring in real industry scenarios as collected and reported by prior work~\cite{huang2022metastable, robloxOutage, gunawi2016whyCloudStop}. The study of previous metastable failures occurring in real-world scenarios reveals that over $50\%$ of these incidents involved retry storms---similar to the example above---as the sustaining artificial overload~\cite{huang2022metastable}. Due to their significance, we focus on retry storms in our discussion and analysis. However, our findings are applicable to a wider range of metastable failures, as we discuss in the rest of the paper.
Another example of a metastable failure is an overload that accumulates background tasks such as garbage collection tasks or compaction tasks. 
The initial accumulation of garbage collection tasks leads to sustaining the overload due to the overhead of garbage collection that leads to a feedback loop of high load and triggering more garbage collection tasks.

Existing solutions like exponential back-off~\cite{exponentialGoogle}, circuit breakers~\cite{circuitBreaker2018nygrid}, and LIFO scheduling~\cite{lifoScheduling2011maestro} are used to mitigate work amplification and metastable failures during monitoring. However, these are often tailored to specific failure instances and lack a general approach for addressing metastable failures~\cite{bronson2021metastable, gunawi2016whyCloudStop}.


Exponential back-off, for example, is commonly applied in retry storms but only provides temporary relief without addressing the root cause~\cite{qian2023viciousCycles}. It may even worsen the situation by delaying the detection of metastable states, as it only extends the interval between retries without adapting to the reduced system capacity. 
This approach falls short in handling general metastable failures, especially those unrelated to retry requests, such as failures due to maintenance tasks like garbage collection or data compaction. These scenarios require a solution that can manage shifts in load and capacity irrespective of the cause. 
Our experiments in Section~\ref{sec:reproduction} demonstrate that exponential back-off alone is not sufficient for overcoming metastable failures.

\cat{Metastability in Replicated Storage Systems}
In this work, we focus on understanding metastable failures in replicated storage systems. 
Given their prevalence in production environments, replicated storage systems are frequently exposed to retry storms.
Past studies have identified this type of failure and analyzed related system outages~\cite{bronson2021metastable, huang2022metastable} but have fallen short of providing a formal analytical model for these failures. We build upon prior efforts by offering an analytical model of metastable failures and enhancing our understanding of them. 
%


We propose \ourModel, an analytical framework to model metastable failures that integrates and extends modeling tools, including queuing theory~\cite{kleinrock1975queueing}, Markov Chains~\cite{norris1998markovchain}, Monte Carlo Analysis~\cite{robert1999monte}, as well as new analytical methods that we develop such as \emph{distance divergence} and \emph{orbit space}. 
The nature of metastable failures is challenging to model analytically and requires us to build a multi-faceted model to describe a range of behaviors and states of the system before, during, and after triggering events. For each stage, the queuing model component captures the behavior of the system in terms of load on the system and the ensuing overload in terms of the accumulation of requests (in the queue) and the accumulation of retry requests. To model retry requests, we introduce a component in queuing models called an \emph{orbit space}. The orbit space models the requests that are being retried.

A limitation of queuing models that we overcome is that a queuing model can be used to analyze the state of the model for only a single instance of the model with a specific configuration and workload~\cite{shortle2018fundamentals}. However, in metastable failures, we need to model the behavior of the system across different stages of different workload characteristics (i.e., before, during, and after the triggering event). This is important to enable capturing whether the sustained artificial overload continues after the incoming traffic and workload return to normal. To this end, we propose a new analytical method called \emph{distance divergence} that can model the overload of the system given a certain configuration as well as a prior state of the configuration. Distance divergence helps us understand if a certain set of configurations and triggering event characteristics lead to a metastable failure.

We perform real experiments that show and validate the accuracy of \ourModel. The experimental validation is performed on a real replicated storage system that utilizes PostgreSQL and Paxos. For each validation run, we measure the expected behavior of the system using the \ourModel, and then perform an experiment with the same parameters. We then compare the predictions of \ourModel\ with the real outcome from experiments. We show that  \ourModel\ closely matches the corresponding real experiments.


The main contribution of our work is proposing a queuing-based analytical model to study and understand metastable failures. Our model proposes new concepts---such as orbit space and distance divergence---that allow us to model metastable failures that were intractable to model prior to our work. We also validate the accuracy of \ourModel\ with real experiments.
Due to space constraints, in our presentation and experimental validation, we focus on replicated storage systems. However, we also provide a description of how \ourModel\ can be utilized and extended for various other use cases, including more complex systems with caches and heterogeneous components.



The remainder of this paper is organized as follows: Section~\ref{sec:background} provides background information on metastable failures and the analytical tools we use. Section~\ref{sec:ms_in_replication_systems} explores and reproduces various metastable failures in a consensus-based replicated storage system. In Section~\ref{sec:queuing-framework}, we introduce the design of \ourModel\ that includes the queuing model, Markov Chain, Monte Carlo analysis, and distance divergence. In Section~\ref{sec:validation}, we present real experiments to validate \ourModel's accuracy. Section~\ref{sec:related_works} reviews relevant literature, and finally, we conclude the paper in Section~\ref{sec:conclusion}.

\section{Background}
\label{sec:background}







\subsection{Metastable Failures}
Metastable failures in distributed systems, formalized recently by~\cite{bronson2021metastable}, are a unique pattern of failures that occur frequently in real industry systems \cite{robloxOutage, huang2022metastable}. They often emerge from optimizations and policies implemented to enhance system performance. However, under certain conditions, these changes can inadvertently trigger a negative impact. For instance, request retries, though aimed at improving reliability, have been identified as a key contributor to metastable failures~\cite{huang2022metastable}.

\cat{Life Cycle}
Figure~\ref{fig:ms-failure-lifecycle} shows the life cycle of a metastable failure, consisting of three stages: stable, vulnerable, and metastable. Initially, the system functions normally without significant load. A change in this load can transition the system into a vulnerable state, where it remains operational and not overloaded but at risk of entering a challenging-to-recover metastable state due to a \emph{triggering event}. The triggering event is an event that pushes the system past a certain load threshold or diminish the system's capacity, e.g., a sudden sharp increase in incoming traffic or a failure of a cache. 

Even after eliminating the triggering event, the system may persist in a metastable state. 
This is due to a \emph{sustaining effect} that creates a feedback loop that persists the overload on the system, e.g., retry requests. 
The metastable state is considered a failure because the system experiences an extremely low goodput even after removing the triggering event.



\cat{Metastable Failure Scenario}
Consider a database with a cache that has $90\%$ cache hit ratio (Figure~\ref{fig:ms-timeline}).
Suppose that, without the cache, the database can process a throughput of $300$ requests/second.
Introducing the cache allows the whole system---which represents the database and the cache---to process $3000$ requests/second. A typical deployment decision would be to maximize the throughput of the whole system, which may lead to allocating around $3000$ requests/second. In such a case, a crash-failure of the cache may lead to the following metastable failure. 
The cache fails, reducing the capacity of the system to $300$ requests/second. The system---without a cache---can only handle $300$ requests/second and is not able to process the continuing flow of $3000$ requests/second. This leads to delaying and dropping requests. These delayed and dropped requests lead to their application clients sending \textsf{retry} requests. These \textsf{retry} requests increase the load on the system to $6000$, which is higher than the client workload of $3000$ ($6000$ is the sum of the incoming load $3000$ in addition to the additional load from retry requests.)  

Eventually, when the cache crash-failure is fixed, the capacity of the system is back to $3000$ requests/s. However, at that point, the load on the system is $6000$ requests/s and thus cannot be handled even with the cache that has a capacity of 3000 requests/s. This leads to sustaining the impact of the crash-failure on performance as requests are still delayed/dropped, and \textsf{retry} requests are still being generated at high rates---a \emph{retry storm}. These retry requests represent the sustaining artificial workload.

 Metastable failures can lead to widespread outages that can disrupt services, potentially lasting from minutes to hours~\cite{robloxOutage}. Unlike logic bugs, metastable failures represent emergent behaviors, meaning they are often not detectable through standard unit or integration tests~\cite{bronson2021metastable}. 

\subsection{Queuing Theory}

Queuing theory is a mathematical approach to the study of waiting lines or queues~\cite{kleinrock1975queueing}. It is used to predict variables like wait times and queue lengths~\cite{shortle2018fundamentals}. 
%
In this section, we provide preliminaries about queuing theory that we utilize in the design of \ourModel.

\cat{$\mathbf{M/M/1/\infty}$ Model}
In a queuing model, requests arrive to the queue, wait for their turn while previous requests are being processed, undergo processing for a certain duration, and then depart from the queue. There are various types of queue models depending on the distribution of workload, processing, number and size of queues, and other characteristics.
A foundational model in queuing theory is the $M/M/1/\infty$ queue (Figure~\ref{fig:mm1}). This model represents a system where arrivals, or rate of requests, follow a Markovian (Poisson) process denoted by the first ``$M$''.
Service times, or the time taken to process requests once at the front of the queue, also follow a Markovian (exponential) distribution represented by the second ``$M$''. 
The model assumes a single server, represented by ``$1$'', to process these requests. Finally, it assumes an infinite buffer space for customers waiting in the queue, denoted by ``$\infty$''.
This model is widely used due to its mathematical tractability and because it can effectively approximate many real-world systems.

A unique feature of the Markovian process is its' \emph{memorylessness}, meaning that the system's transition from its current state to the next is only dependent on the current state and is not influenced by any past events. Thus, in a queuing system, each request arrives independently from others, and the time spent processing a request is unrelated to previous events.  
An additional characteristic of a Poisson process, known as the superposition principle~\cite{kingman1992poisson}, allows the combination of two or more independent Poisson processes to form a single, unified Poisson process.
These properties make the analysis of the queuing system tractable.

We refer to a queuing system as ``Stationary'' when its key characteristics, like the average number of waiting requests or the arrival rate, remain constant over time. In simpler terms, a system reaches a stationary state when its inflow and outflow rates are in equilibrium, leading to a time-independent probability distribution of the state of the system~\cite{BrandtPrankenLisek1987stationryqueue}.

\cat{Markov Chain}   
A Markov chain~\cite{markov1971markov} is a mathematical process that is represented as a state-transition diagram, with the probability of each transition depending only on the current state and not on the path that led to it (i.e., it is memoryless). It is particularly useful in the analysis of queuing systems, as many queuing models can be described as Markov processes.
A Continuous-Time Markov Chain (CTMC)~\cite{anderson2012ctmc} is a stochastic model used to describe systems transitioning between states over continuous time, governed by the Markov property. Each state transition in a CTMC is associated with a certain rate, indicating how rapidly the system is likely to transition. The time spent in one state before transitioning to the next typically follows an exponential distribution.

In a queuing system, each state represents a different processing stage, i.e., how many requests are being processed/queued. For instance, one state might represent two requests being serviced and three requests waiting.
A CTMC can model a queuing system's behavior, where transition rates symbolize the rates at which requests arrive (arrival rate) or get serviced (service rate). 

A CTMC is termed ``Stationary'' if transition probabilities depend only on the time span between transitions rather than absolute time. This means the process behaves the same at all times, and each state's probability distribution is measurable.

\cat{Monte Carlo Analysis}
Monte Carlo analysis~\cite{robert1999monte, mahadevan1997monte} offers a powerful tool to calculate the approximate probability of being in each state in CTMC. This analysis involves generating random samples and numerically evaluating them repeatedly. This analysis is especially useful when deriving exact state probabilities is intractable. 


\cat{Retrial Queues}
Retrial requests frequently arises in many queuing systems, which led to the development of retrial queues~\cite{phung2019retrialsurvey}. In these systems, if requests don't get processed immediately or if they're blocked, they're placed in a virtual ``waiting room'', often referred to as the \emph{orbit queue}. Requests residing in the orbit queue are retried according to the system's retrial policy. Under a classic retrial policy, the time a request spends in the orbit is randomly determined, following an exponential distribution, and each request is retried independently of others. 
The traditional retrial queues are limiting as they only place requests into the orbit when blocked, which does not accurately reflect many real systems' retrials.  In the paper, we extend this notion of a retial queue to our proposed \emph{retrial orbit} that can more accurately model retrials.






\section{Metastable failures in Replication Systems}
\label{sec:ms_in_replication_systems}
In this section, we discuss various instances of metastable failures that may occur in a cluster of replicated storage systems. We study and reproduce these failures within a consensus-based replication system that runs a PostgresSQL database. The aim of this section is to provide intuition of how metastable failures occur to inform the design of \ourModel.

\subsection{System Model}
\label{sec:system_model}
The system consists of a replication group with $n=3$ nodes and $m=n=3$ data partitions (Call the three nodes $A$, $B$, and $C$ and the three partitions $p_a$, $p_b$, and $p_c$.)
We pick this cluster size as it is typical in industry and for ease of exposition. The outcomes of this section apply to clusters of other sizes as well.
 Each partition's state is maintained via a State-Machine Replication (SMR) log with a consensus-based replication protocol~\cite{du2009multi, lamport2001paxos, mazieres2007paxosPractical}. To balance the load across the three nodes, each node is assigned to be the leader of one of the partitions. 

\cat{Data Model}
\label{sec:data_model}
Each data partition is a key-value store with a mutually exclusive set of keys. Clients issue database transactions that consist of read and write operations to be processed according to ACID and serializability guarantees~\cite{gray1992transaction}.
We assume that workload generation is independent of the status of the system. Specifically, clients of each partition generate $W$ transactions/second, where a transaction consists of a group of read and write operations. 

\cat{System Failures}
To distinguish the different types of failures, we will use the terms \emph{crash-failure} and \emph{metastable failure} throughout the document. A crash-failure refers to a scenario where one or more machines in the system cease to function. In contrast, a metastable failure influences the entire system, as elaborated in Section~\ref{sec:background}.
Up to $f=1$ machines can crash at any point in time. Data is persisted on disk so that after a failure, the state of the node can be recovered. Communication messages can be reordered and/or indefinitely delayed---i.e., we assume an asynchronous communication model~\cite{charron1996synchronous}.

\subsection{Metastable Failure Scenarios}
In this section, we outline scenarios in which a metastable failure can occur due to retry requests within a replicated storage system.


\cat{Scenario 1 (Node Failure)}
Consider a crash-failure of node $A$ during normal-case operation. In this case, transactions to partition $p_a$ can no longer be processed until another node takes over as the leader of $p_a$. In such a case, another node---for example $B$---takes over and becomes the leader of both $p_a$ and $p_b$. This increases the load on $B$ as it is managing the state of both partitions. This may lead to overloading $B$, delaying responses to transactions, and even dropping requests.

There are two outcomes of the delaying and dropping of requests: (1)~a buildup in the pool of requests occurs. The amount of build-up is proportional to the duration of the crash-failure and return to the normal-case operation. Also, it is proportional to the difference between the capacity of the replication system before and after the crash-failure. (2)~Application clients will start issuing \textit{retry} requests due to the delayed responses. We define a \textit{retry} request as the following: if no response is received to a request within $\tau$ time, then a \textit{retry} request is sent. Then, if no response is received after sending the \textit{retry} request within $\tau$ time, then another \textit{retry} request is sent. This continues until a response is received for the corresponding transaction or the number of retrials reaches a predefined limit.

After node $A$ recovers, it reclaims the leadership of partition $p_a$. After restoring the normal-case state, there is an impact from the failure in the buildup of the pool of requests as well as the \textit{retry} requests which place the system into a metastable state. 
This is because retry requests are continuously being generated even after recovery from the crash failure due to the buildup and feedback loop of retry requests that delays both current and future requests.

\cat{Scenario 2 (Load Surge)}
\label{sec:load-surge}
A simpler metastable failure operation in a replicated storage system is due to a load surge. Consider a case of load surge due to a triggering event---for example, increased traffic during a holiday~\cite{robloxOutage}. If the system is operating in a vulnerable state, this event may lead to overloading the system. 
This results in the issuance of \textit{retry} requests due to delayed responses. Even after the trigger is removed and the incoming load returns to normal, the system remains in an overloaded state due to the backlog of \textit{retry} requests. This feedback loop of continuous \textit{retry} requests persists the overload, which makes recovery difficult.

The previous two examples demonstrate that retry requests play a critical role in perpetuating metastable failures (previous studies indicate that more than 50\% of metastable failures in practice are due to retry storms \cite{huang2022metastable}.) This is because retry storms can be triggered in various unpredictable ways and combinations, making remedies and anti-patterns unable to solve the problem of retry storms in general. By focusing on retry storms, we can address and mitigate one of the common issues associated with metastable failures. 

In the following sections, we will focus on the ``Load Surge'' scenario to regenerate and investigate metastable failures in a replicated storage system.

\subsection{Formalizing Metastable Failure}

In this section, we provide a formalization of metastable failures that build on and extend prior work formalism~\cite{huang2022metastable} to introduce specific formalization for metastable failures in replicated storage systems. 

\begin{figure*}[t]
  \centering

    \begin{minipage}[b]{0.31\textwidth}
    \centering
    \includegraphics[width=\linewidth]{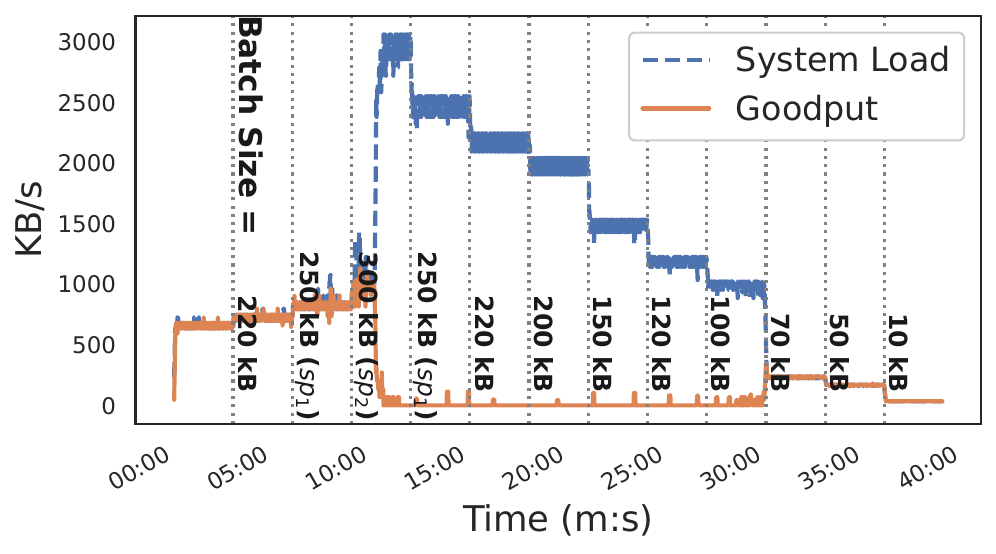}
	\subcaption{Goodput}
	\label{fig:batch-goodput}
  \end{minipage}
   \hfill
  \begin{minipage}[b]{0.33\textwidth}
    \centering
    \includegraphics[width=\linewidth]{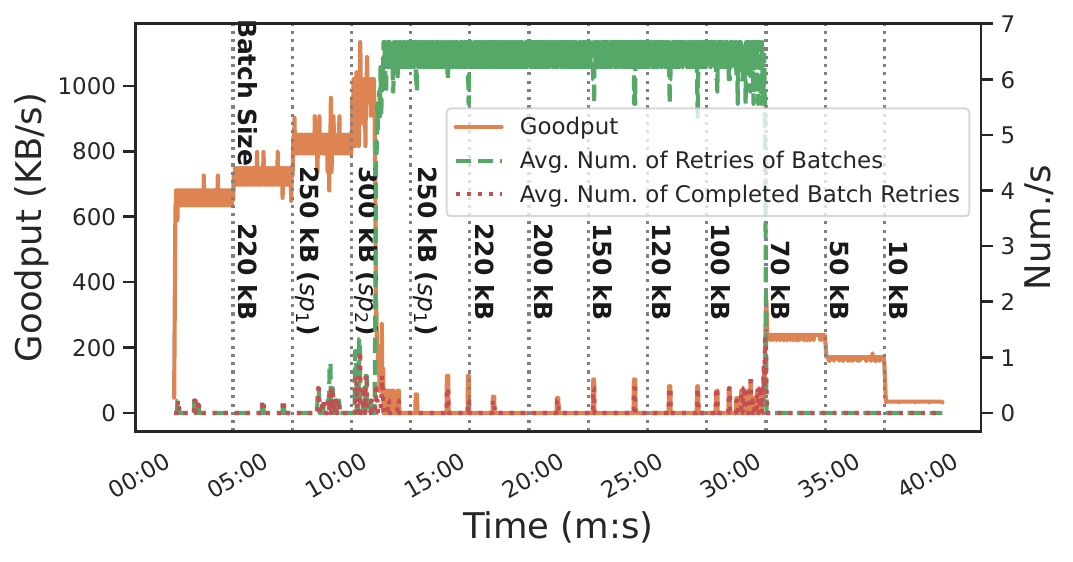}
    \subcaption{Average number of retrials}
    \label{fig:batch-retry}
  \end{minipage}
  \hfill
  \begin{minipage}[b]{0.345\textwidth}
    \centering
    \includegraphics[width=\linewidth]{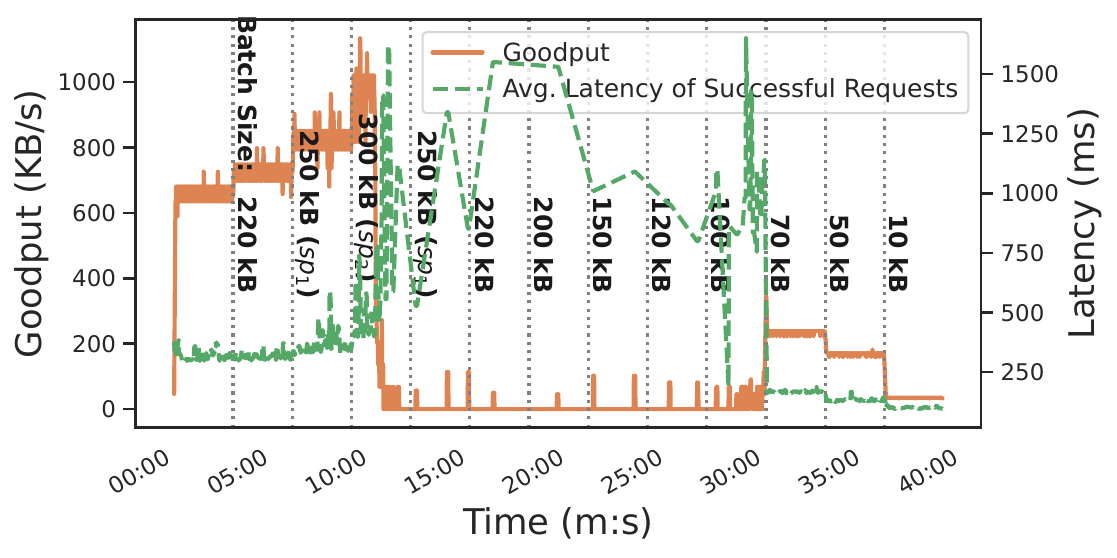}
    \subcaption{Average latency of successful transactions}
    \label{fig:batch-latency}
  \end{minipage}
  \caption{The impact of varying the batch size on goodput, retrials, and latency in a metastable failure scenario.}
\end{figure*}

\cat{The scaling parameter}
\label{sec:scaling-parameter}
We first define the notion of a \emph{scaling parameter} (SP). The scaling parameter is a system configuration parameter in the system that controls the utilization of the system. Typically, such scaling parameters are configured to maximize the utilization of the system's resources.
In this paper, we consider the \emph{batch size} and/or \emph{incoming throughput} as the SP. Batching is a widely-used technique to allow utilizing the resources more efficiently by grouping tasks to be processed together rather than individually. In the replication system that we consider, batching is implemented by assembling transactions into groups and subsequently committing them to the SMR log. Each batch constitutes an entry in the SMR log, which effectively reduces the communication required per transaction, resulting in improved resource utilization.

Changing the SP leads to changing performance metrics of interest. The three performance metrics we consider are (1) system load---the number of transactions incoming to the system, including retry requests and client requests, (2) goodput---the number of processed transactions per second; and (3)~latency---the average response time to a user request.

In the ``Load Surge'' scenario, the system's resources are not fully utilized when the batch size is initially low. As we start increasing the batch size, the system can accommodate the additional workload. This stage is referred to as the \emph{scaling} phase, wherein increasing the batch size (SP) leads to a near-linear performance (goodput) improvement.
However, the system reaches a vulnerable state at a certain batch size ($sp_1$). Further increasing the load on the system may cause it to enter a metastable state. Eventually, at another batch size ($sp_2$), the system starts experiencing retry requests that force it into a metastable state, leading to substantial delays or dropped requests. Even reducing the batch size to $sp_1$ is not enough to recover the system requiring a costly recovery operation that significantly reduces the system load.

\cat{The Metastable Recovery Fallacy}
We now describe a fallacy in how systems behave after recovery.
Here, we use the term recovery to denote the case of a system load (e.g., incoming traffic) coming back to the normal load after the trigger.
Consider a system that is configured with batch size $sp_i$ (where $sp_i<sp_2$) and achieves a goodput value of $X$. Consider the case that a load is triggered to an overloaded value of $sp_2$ and then, after some time, the load recovers back to $sp_i$. Let's call this state the \emph{aftermath} state. At the aftermath state, the question is \emph{would the behavior of the system retain the original performance pattern and values? In other words, does the aftermath state differ from a metastable state?} 

The analysis and experience reported in the previous metastable failure papers~\cite{bronson2021metastable, huang2022metastable} show that the return to the same performance pattern and values is a fallacy---as we describe in earlier discussions. Rather, what happens is that the load on the system becomes much higher than expected because of retry requests and the goodput value is $Y$ which is much lower than $X$. The system is in a metastable state.

\subsection{Reproducing a Metastable Failure} 
\label{sec:reproduction}
To substantiate our prior discussion on metastable failures, this section centers around experiments conducted to reproduce the ``Load Surge'' scenario in a replication system. The goal is to study and showcase metastable behavior with real experiments. We begin by detailing the architecture and setup of the utilized prototype, followed by an exploration of experiments that induce a load surge metastable failure in a replication system.

\cat{Prototype} 
\label{sec:prototype}
Our prototype employs a Java-based implementation of the multi-paxos protocol~\cite{du2009multi}. Each Paxos instance is integrated with PostgreSQL for data storage. The prototype accommodates transaction batching, allowing clients to batch-send their requests. Partitioning is also incorporated, with each partition running a distinct Paxos instance for replication across other servers. The prototype follows the data model presented in Section~\ref{sec:data_model}. We have used an asynchronous approach for inter-node communication, implemented via the gRPC protocol~\cite{indrasiri2020grpc}.

\cat{Setup}
We deployed three servers on a distributed setup using the Chameleon Cloud platform~\cite{keahey2020chameleon} utilizing its Compute Haswell nodes. 
One server, designated as the leader, runs the prototype to accept client requests and replicate data across the other two servers, subsequently confirming successful completion to the client.

Clients operating from another node in the distributed setup send their transactions as batch requests. Each request is dispatched according to a set \textit{interval} time. If a response is not received within a \textit{timeout} period, the request is retried and immediately resent. This retry process is limited to a maximum number of attempts, after which the request is discarded.

\cat{Scenario} 
The system load is incrementally increased until the system reaches a vulnerable state. A sudden surge in incoming load can push the system into an overloaded state, where retry requests could keep the system in the metastable state, resulting in a metastable failure.

We follow Section~\ref{sec:scaling-parameter} formalization and use batch size as the scaling parameter.
We set the interval between each request to $300 ms$, meaning transactions are dispatched in batches every $300 ms$ asynchronously, irrespective of other requests. Each request has a timeout of $600 ms$ and is immediately retried upon timeout. Each request is retried a maximum of $3$ times, including the initial user attempt. We increased the batch size in a step-wise manner, beginning at $220 KB$ and peaking at $300 KB$, which pushes the system into an overloaded state. Thereafter, the batch size is reduced stepwise to a low value of $10 KB$.

In this scenario, having a value of $sp_1 = 250 KB$ leaves the system in a vulnerable state, and a scaling parameter of $sp_2 = 300KB$ pushes the system into the metastable state.


Figure~\ref{fig:batch-goodput} displays the goodput and system load in this experiment. Initially, when the batch size is low (below $250KB$), resources are not fully utilized and responses are timely. As we increase the batch size (SP), system goodput increases correspondingly. However, at a certain batch size ($250KB$), the system is vulnerable, and a load spike could transition the system into a metastable state. An increase in batch size to $300KB$ (at time \textit{09:00}) overloads the system, leading to a metastable state.

Retry requests put even more work on the system, increasing the system load significantly. Decreasing the batch size to its previous value of $250KB$ (at time \textit{12:00}) does not recover the problem, and the system remains overloaded, with a goodput of nearly $0$. Here, a metastable failure has occured, and to recover, the batch size must be significantly reduced to $70KB$ (shown at time \textit{30:00}) to accommodate retry requests. 


Figure~\ref{fig:batch-retry} illustrates the impact of increasing the batch size on the average number of retry requests and their success rate. As indicated, during periods of system failure, retry requests could potentially saturate the system, with most requests requiring retries. These retry requests generate a feedback loop, creating a sustaining artificial load that perpetuates the metastable failure.

\begin{figure}[t]
	\centering
	\includegraphics[width=.85\linewidth]{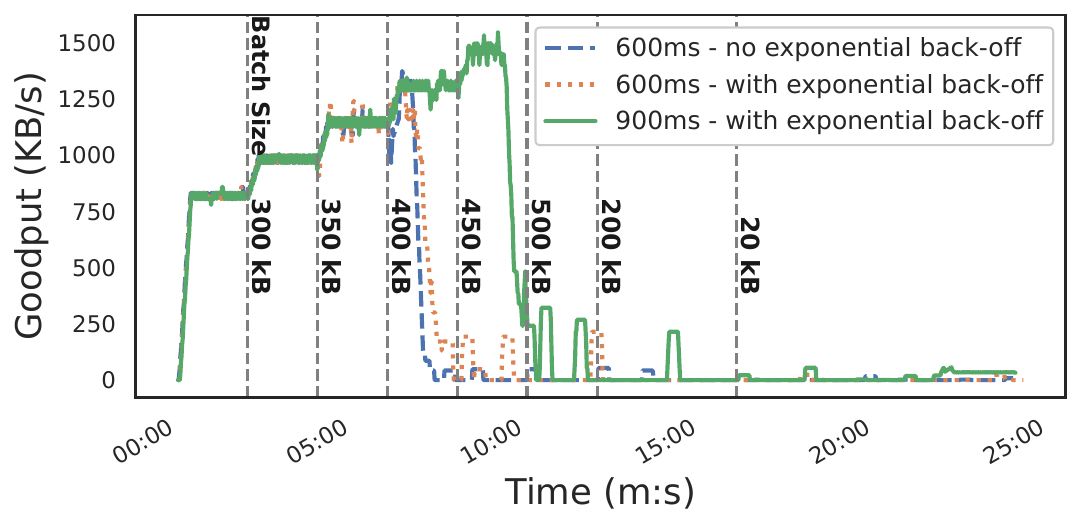}
	\caption{Timeout impact on metastable failure scenarios}
	\label{fig:batch-expnential}
\end{figure}
\begin{figure*}[t]
  \centering

    \begin{minipage}[b]{0.29\textwidth}
    \centering
    \includegraphics[width=\linewidth]{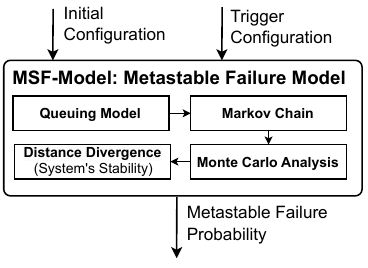}
	\caption{Overview of \ourModelName}
	\label{fig:queuing-framework}
  \end{minipage}
   \hfill
   \begin{minipage}[b]{0.42\textwidth}
    \centering
    \includegraphics[width=\linewidth]{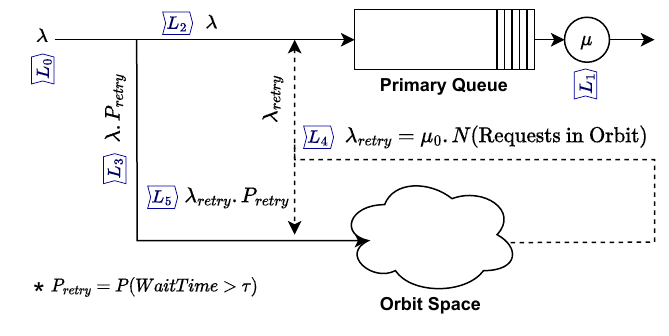}
    \caption{Proposed queuing model's overview}
    \label{fig:queuing-overview}
  \end{minipage}
  \hfill
  \begin{minipage}[b]{0.27\textwidth}
    \centering
    \includegraphics[width=\linewidth]{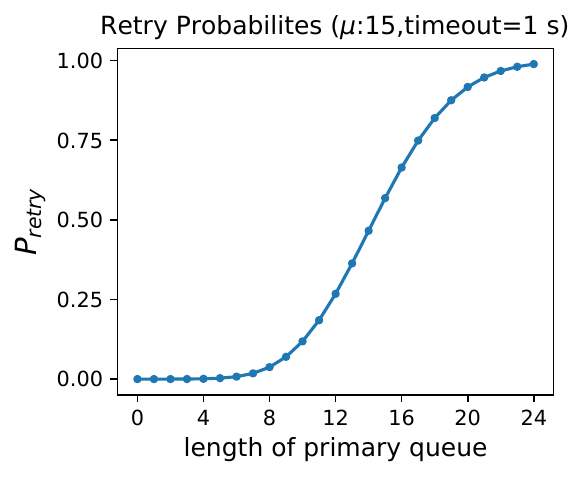}
    \caption{Retry probabilities example}
    \label{fig:retry_probs}
  \end{minipage}
\end{figure*}

Figure~\ref{fig:batch-latency} shows the average latency of successful requests. Initially, with smaller batch sizes, the latency is less than $400ms$. However, as the batch size grows, latency not only increases but also becomes more variable. 
When the system is in the metastable state, most requests fail, and the few that succeed experience significantly increased latency.

\cat{Exponential Back-Off}
We integrated exponential back-off into our prototype for a comparative analysis of metastable failures when this strategy is employed by the client. To better observe the effects of exponential back-off, we increased the maximum number of retries to 4. As shown in Figure~\ref{fig:batch-expnential}, we observed three distinct metastable failures at three varying initial timeout values. The results show that even with a 600ms timeout and exponential back-off, metastable failures still occur, similar to the scenario without exponential back-off. When the timeout is increased to 900ms with exponential back-off, the onset of metastable failure is delayed, occurring at a higher trigger point.

\section{A Queuing Framework of Retry Storms}
\label{sec:queuing-framework}

In this section, we propose the design of \ourModel, a queuing framework for requests and retry request buildup for a replicated storage system. The goal of \ourModel\ is to capture the behavior of the replicated storage system and build an analytical framework for analyzing metastable failures.

We begin by defining the queuing model, and then describe how to build and analyze the corresponding Markov process. The Markov process allows us to compute the probability of different states of the system based on the input parameters (such as the workload and processing rate). Then, we utilize the Markov process to build a framework for modeling metastable failures, called \ourModel. This metastable failure framework can be used to answer whether a certain configuration/workload of the system would lead to a vulnerable and/or metastable state. The framework works by deriving the workload buildup that can be incurred from a Markov process configuration (that corresponds to a triggered event in a vulnerable state). It then determines whether this buildup would lead to sustaining the failure (after returning to the normal configuration). 
Finally, we conclude the section with a discussion of how \ourModel\ can be used and extended for various use cases.
Figure~\ref{fig:queuing-framework} provides a visual representation of the components of our proposed queuing framework.


\subsection{Queuing Model}

\cat{Definition}
Figure~\ref{fig:queuing-overview} presents an overview of the proposed queuing model. This model consists of an $M/M/1/\infty$ primary queue and a secondary orbit space to model retry requests.
Requests in this model follow a Poisson process with an arrival rate of $\lambda$ ($L_0$ in Figure~\ref{fig:queuing-overview}), while the service time for each request is exponentially distributed with a processing rate of $\mu$ ($L_1$). The $M/M/1/\infty$ primary queue models the replicated storage introduced in Section~\ref{sec:system_model}, which sequentially receives and processes requests.


We observe that modeling metastable failures requires some ``memory'' since they are impacted by a sustaining effect, followed by a triggering event during a vulnerable state. This ``memory'' characteristic needs to be captured by the queuing model.
Given that the primary queue, due to its memoryless property, does not directly model a timeout, an orbit space is introduced. This orbit space serves as the ``memory'' component, enabling the capture of timeouts and retries that contribute to metastable failures. The orbit space can be thought of as a virtual waiting area where requests reside after failing to receive service during their initial attempt within the configured timeout and are waiting to retry.

The probability of retrying requests ($P_{retry}$) is determined by the wait time of the request exceeding a specific timeout ($P_{retry} = P(\text{Wait Time} > \tau)$). Based on this probability, requests enter the primary queue ($L_2$) and are potentially added to the orbit space ($L_3$). Here the orbit space captures the requests that are timed out.

Request in the orbit space reattempt service following a Poisson process with a rate of $\mu_0 = \frac{1}{\tau}$. This is based on the assumption that customers will retry their requests after the configured timeout.
Therefore, the overall incoming rate of requests from the orbit space is equivalent to $\lambda_{retry} = N.\mu_0$, with $N$ representing the number of requests in the orbit space ($L_4$). 
This is because each request in the orbit space is retried independently, regardless of the status of other requests in the orbit space.

The model also accounts for the potential of re-retrying requests in orbit space. Depending on the probability of retry, each of the requests in the orbit space will be added to the primary queue, and potentially another retry request will be added to the orbit space ($L_5$).






\cat{Probability of Retry}
$P_{\text{retry}}$ is a function of the queue length, processing rate, and timeout ($P_{\text{retry}} = F(\tau, L, \mu) $). We will use the notation $P_i$ to denote the retry probability with $i$ requests in the queue; e.g., $P_1$ is equivalent to $P_{\text{retry}}$ when one request is waiting to be serviced in the primary queue. 



Figure~\ref{fig:retry_probs} shows a plot of an example calculation of retry probabilities based on the length of the primary queue, assuming $\mu=15$ and $\tau=1s$. It is important to note that a higher $P_{\text{retry}}$ adds more packets to the orbit space and consequently increases the primary queue length after a retry, making the retry probability even higher. This feedback loop models the sustaining effect in a metastable failure following a triggering event. The figure shows that a metastable failure can easily occur if a trigger rapidly increases the primary queue length. In this experiment, once the queue length reaches $24$, the probability of a retry becomes one, meaning all subsequent requests will be retried.

\subsection{Markov Chain}
\textbf{Definition.}
We propose a two-dimensional continuous-time Markov chain (CTMC)~\cite{markov1971markov, norris1998markovchain} model to analyze the performance of our proposed queuing model. 
This Markov chain is composed of the states of form $(i, j)$, where $i$ denotes the number of requests in the orbit space, and $j$ represents the number of requests in the primary queue waiting to be serviced.

\begin{figure}[t]
	\centering
	\includegraphics[width=0.9\linewidth]{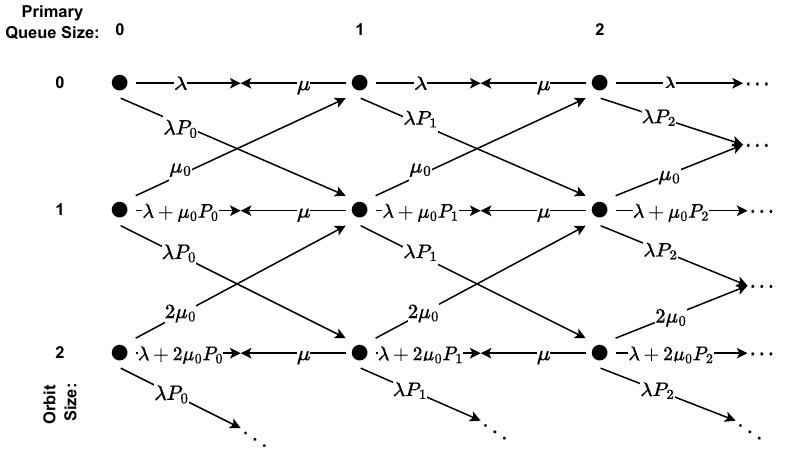}
	\caption{Proposed queuing model's transition diagram}
	\label{fig:queuing-transition}
\end{figure}

To visualize the process, a transition diagram---shown in Figure~\ref{fig:queuing-transition}---is employed, which depicts the possible state transitions and their corresponding rates. In the diagram, each state $(i, j)$ is represented by a node, and the arrows between nodes represent the transition probabilities. Each row in the diagram corresponds to the number of requests in the orbit space, and moving to lower levels indicates an increase in requests within the orbit space. Each column shows the primary queue request count, with columns to the right representing an increased number of requests in the primary queue. The system has three types of transitions:

\begin{enumerate}[leftmargin=*]
    \itemsep0em 
    \item \textbf{Arrival transitions}: A new request arrives at the system and will be added to the primary queue. This transition occurs from state $(i, j)$ to state $(i, j+1)$ at a rate of $\lambda$. 
    If the request is retried, it will also be added to the orbit space, transitioning from state $(i, j)$ to state $(i+1, j+1)$ at a rate of $\lambda P_{j}$.
    \item \textbf{Departure transition}: A request departs from the system, creating room for another request in the primary queue. This transition occurs from the state $(i, j)$ to state $(i, j-1)$ at a rate of $\mu$, where $\mu$ is the service rate.
    \item \textbf{Retrial transitions}: One of the requests in the orbit space gets retried, and will be added to the primary queue. A transition occurs from state $(i, j)$ to $(i-1, j+1)$ at a rate $N\mu_0$, moving the request from orbit space to the primary queue. 
    The request may also be retried and added back to the orbit space, transitioning from $(i, j)$ to $(i, j+1)$ at a rate of $N\mu_0 P_j$.
\end{enumerate}

    


The Poisson process superposition principle~\cite{kingman1992poisson} allows us to consider transitions with the same state change as a single transition at a rate equal to the sum of all transition rates. In the following sections, we use this two-dimensional CTMC model for a deeper understanding of the underlying dynamics of the system.

The result of a stable Markov chain is its long-term behavior, which is described by the unique stationary distribution. The stationary distribution is a probability distribution over the states of the Markov chain that remains unchanged as the process evolves over time. When a Markov chain is stable, it is guaranteed to converge to this stationary distribution, regardless of the initial state. In other words, after a sufficiently large number of transitions, the probabilities of being in each state of the Markov chain will approach the values given by the stationary distribution. 
In the following, we analyze the Markov chain to derive the stationary distribution that allows us to have a better understanding of the queuing model.


\cat{Monte Carlo Analysis}
We utilize the Monte Carlo simulation\cite{robert1999monte} method for estimating the stationary probabilities of the continuous-time Markov chain (CTMC) in the context of our proposed queuing model.

\subsection{Distance Divergence}
In this section, we propose the notion of \emph{distance divergence} that aims to measure the stability of the system. At a high level, a non-stationary system is one where the queues tend to grow and not stabilize within certain sizes. The distance divergence is a measure of this high-level behavior.
In a stationary system, a time-independent probability distribution exists for the system's state space. However, in a non-stationary system, such a distribution doesn't exist.

\cat{Distance Metric}
We propose a metric called \emph{distance metric} to represent the system's stability with a constant and converging number for a stationary system, and an increasing value for a non-stationary system derived from the probability distribution of the system's state space.

This metric is calculated by multiplying the distance of each state from the $(0, 0)$ state with the probability of being in that state. Therefore, states further from the $(0, 0)$ state, indicating more requests in the primary queue and orbit space, carry more weight and contribute more significantly to the distance metric.

\fontsize{8}{8}\selectfont
\begin{equation}
   \text{Distance Metric} = \sum_{(i, j) = (0, 0)}^{(i, j) \in \text{states}} \sqrt{i^2 + j^2} \times P_{ij} 
\end{equation}
\normalsize

\begin{figure}[t]
     \centering
       \begin{minipage}[b]{0.48\linewidth}
 \centering

    \includegraphics[width=\linewidth]{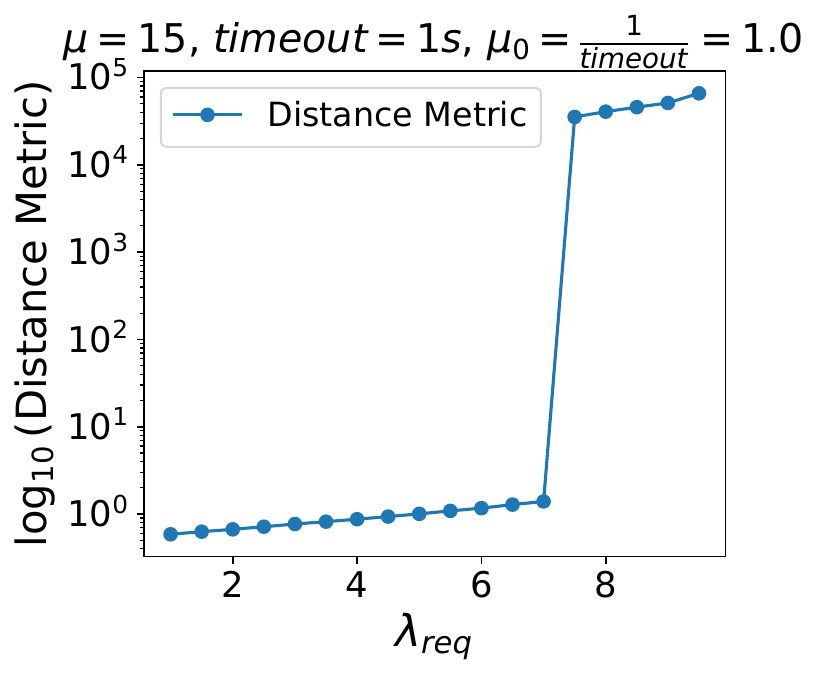}
    \caption{Distance Metric as a function of $\lambda$ ($\log_{10}$ scale)}
    \label{fig:distance_metric}
  \end{minipage}
    \hfill
  \begin{minipage}[b]{0.46\linewidth}
    \centering
    \includegraphics[width=\linewidth]{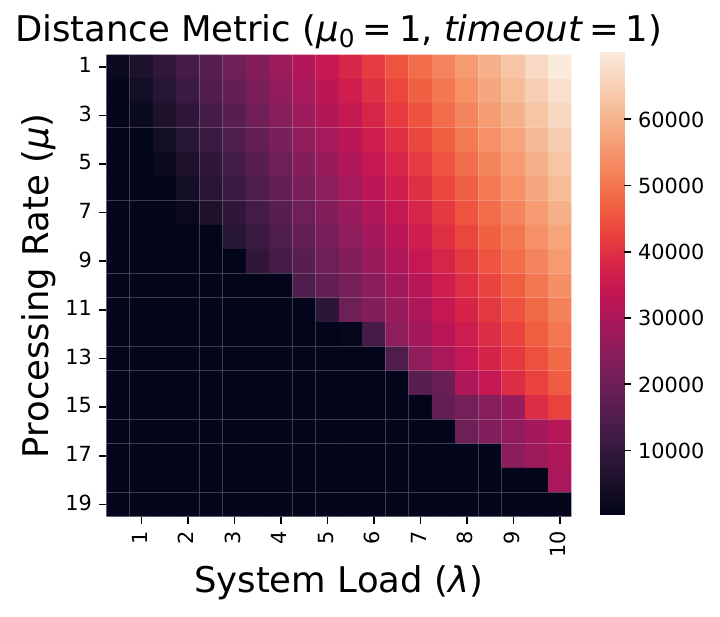}
    \caption{Values of Distance Metric for different $\lambda$ and $\mu$}
    \label{fig:distance_metric_lambda_mu}
  \end{minipage}
  \hfill

\end{figure}

Here $P_{ij}$ shows the stationary probability for the state $(i, j)$. Stationary probabilities can be computed using the Monte Carlo simulation introduced in the previous section. 

In a stationary system, after several iterations of the simulation, the \textit{distance metric} converges to a specific value due to the convergence of probabilities. Conversely, in an unstable system, this metric continuously increases as requests accumulate in the primary queue and orbit space.

Figure~\ref{fig:distance_metric} represents the distance metric value as we increase the arrival rate, $\lambda$. In this experiment, we maintain a constant processing rate, $\mu$, at $15$. Each request has a timeout set at $1$ second, which makes the retry rate, $\mu_0$, equivalent to $1$. Figure~\ref{fig:distance_metric} illustrates that as $\lambda$ increases, the probability of having requests in the primary queue and orbit space also increases. This behavior is attributed to the rapid growth of $P_{retry}$ as the length of the primary queue expands. Beyond a certain threshold of $\lambda$, the system becomes non-stationary, causing the distance metric to fail to converge to a specific value and consequently become an exceedingly large number. This experiment shows the difference in the distance metric between a stationary and non-stationary system.

Figure~\ref{fig:distance_metric_lambda_mu} illustrates the variation in the distance metric as both $\lambda$ and $\mu$ parameters are changed. The proximity of the distance metric value to zero represents a stationary system wherein the probability of being in each state can be calculated. This experiment highlights the existence of a boundary beyond which the system ceases to be stationary (colorful blocks). As the processing rate increases, the incoming load that the system can withstand also grows until a certain point at which the system becomes unstable and transitions into a non-stationary system. 

\begin{figure}[t]
  \centering

  \begin{subfigure}[b]{0.50\linewidth}
    \centering
    \includegraphics[width=\linewidth]{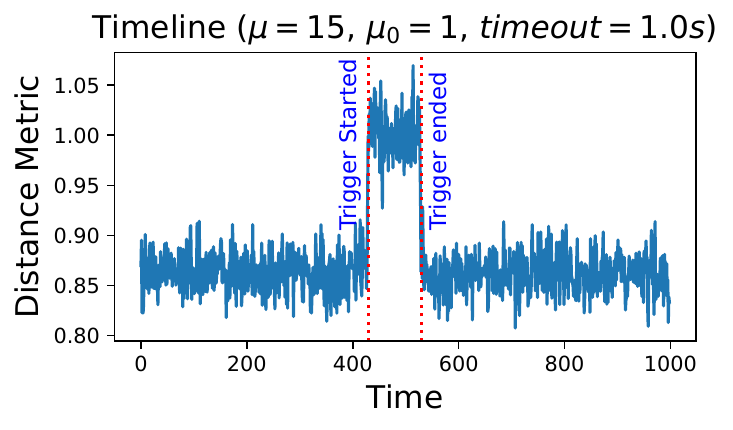}
    \subcaption{Load Spike $\lambda=4$ to $\lambda=5$}
    \label{fig:dm_timeline_nonmeta}
  \end{subfigure}
  \hfill
  \begin{subfigure}[b]{0.48\linewidth}
    \centering
    \includegraphics[width=\linewidth]{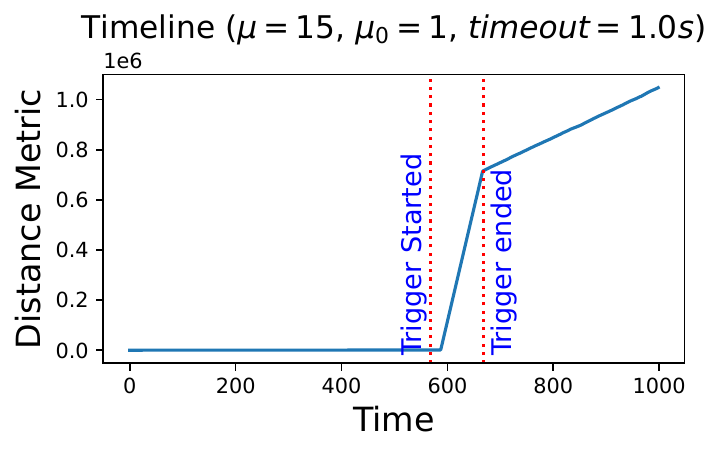}
    \subcaption{Load Spike $\lambda=4$ to $\lambda=6$}
    \label{fig:dm_timeline_meta}
  \end{subfigure}
  \caption{Simulation Timeline}

\end{figure}

A metastable failure timeline can be simulated using our proposed queuing model in conjunction with the distance metric. Figures~\ref{fig:dm_timeline_nonmeta} and~ \ref{fig:dm_timeline_meta} represent timelines for two distinct scenarios. In each instance, simulations are executed over 1000-second intervals with a constant processing rate of $\mu=15$ and a retry rate of $\mu_0=1$. 
In these experiments, we reproduce a load surge scenario, as detailed in Section~\ref{sec:load-surge}, by adjusting the arrival rate as a scaling parameter. 

In Figure~\ref{fig:dm_timeline_nonmeta}, a load spike triggering event elevates the incoming rate as the scaling parameter from $\lambda=4$ ($sp_1$) to $\lambda=5$ ($sp_2$) for a duration of $100$ intervals. Upon recovery from the triggering event, the system reverts to its initial state at $\lambda=4$. This experiment shows that the triggering event does not induce a metastable state in the system, and there will not be a sustaining effect. The system remains stationary both before and during the triggering event. After recovery of the triggering event, the system resumes its normal operations, and the distance metric reverts to its former value. 


On the contrary, Figure~\ref{fig:dm_timeline_meta} represents a metastable failure. In this scenario, a load spike elevates $\lambda$ from $4$ ($sp_1$) to $6$ ($sp_2$) for $100$ intervals, while other configurations remain similar to Figure~\ref{fig:dm_timeline_nonmeta}. This experiment reveals that the load spike trigger propels the system into a non-stationary state, leading to a backlog of requests in both the primary queue and orbit space. Following recovery from the load spike, the abundance of requests in the primary queue and orbit space results in a retry probability of one, causing all packets to be retried and acting as a sustaining effect. Consequently, the system persists in a metastable state.
To recover from this metastable failure, it's necessary to either significantly raise the processing rate or discard requests from the queue and orbit space. 


\subsection{Metastable Failure Model}

The primary application of the proposed queuing model and its Markov chains is to enable modeling a metastable failure. In this section, we provide a model of metastable failure, \ourModel, that can be used to derive whether a metastable failure will happen given two sets of input parameters: (1) the configuration of the system in the normal-case scenario and (2) the configuration of the system during the triggering event and its duration. The configuration of both states includes the following parameters: arrival rate, processing rate, retrial rate, as well as request timeout.

Consider a timeline for a replication storage system. Initially, the system processes requests at a rate of $\mu$ and receives requests at a rate of $\lambda$ ($sp_1$). A triggering event may either decrease the system's capacity to $\mu'$ or increase the load on the system to $\lambda'$ for a certain duration of $\Delta_{\text{trigger}}$ ($sp_2$). Both of these events can also occur simultaneously. After this period, the system reverts to its original configuration with the initial parameters.
If the triggering event pushes the system to a metastable state, the system cannot recover and return to normal conditions due to sustaining effects (e.g., retrial requests). There is also a possibility of not entering the metastable state if the trigger's intensity and/or duration are insufficient.

A Monte Carlo simulation can be formalized by its initial state and input configuration. The simulation can be either stationary, providing the probabilities of being in each state, or non-stationary, indicating that the system is unstable.

\fontsize{8}{8}\selectfont
\begin{equation}
    M \left( \left(i, j\right), \mu, \lambda, \mu_0, \tau) \right) 
    =
    \begin{cases}
    \{P_{(i,j)} : i,j \in 	\mathbb{N} \} & \text{stationary} \\
    \varnothing & \text{non-stationary}
    \end{cases}
\end{equation}
\normalsize 

Here, $M$ represents a simulation function with the system's configuration as input parameters. The first parameter ($\left(i, j\right)$) indicates the initial state of the Markov chain, followed by the processing rate ($\mu$), arrival rate ($\lambda$), retrial rate ($\mu_0$), and timeout ($\tau$), respectively. If the simulation is stationary, each state has its own probability, and the distance metric converges to a value; in this case, the function's result is the probability of being in each state. A non-stationary simulation implies that starting from the input state with the specified configuration leads to a continuously growing number of accumulated requests in both the primary queue and orbit space‌, making it impossible to calculate the probability of being in each state. 
This simulation accepts an arbitrary parameter, $\Delta$, to set the simulation time, bypassing the need to check for stopping criteria.

A metastable state is a situation where the system enters a bad state, and this state persists due to a sustaining effect. In the context of a queuing model, this occurs when the queuing system becomes non-stationary. A non-stationary system is unstable, and the existing requests in the orbit space and primary queue amplify the retry probability. This creates an artificial load that keeps the system in its non-stationary state. Therefore, we define a non-stationary queuing system as a system experiencing a metastable failure (the system is in a metastable state).

For a Markov chain, we define non-stationary states as $NS$. Initiating the simulation from these $NS$ states results in a non-stationary simulation.

\fontsize{8}{8}\selectfont
\begin{equation}
     NS(\mu, \lambda, \mu_0, \tau) = 
      \{(i, j):  M\left( \left(i, j\right), \mu , \lambda , \mu_0 , \tau)  \right) = \varnothing , i, j \in 	\mathbb{N}\}
\end{equation}
\normalsize

For a given input configuration, $NS$ is computed by collecting all states such that initiating a Monte Carlo simulation from these states does not lead to a convergence of probabilities for each state. Given that each state is either stationary or non-stationary, we can establish that $S = \overline{NS}$, where $S$ represents all the stationary states.

The timeline of a metastable failure can be simulated. Initially, the system is stable, given the initial configuration. However, a triggering event could transition the system into a metastable state, which can be represented as a state $(i_{MS}, j_{MS})$ in the Markov chain. This state is non-stationary, and beginning from this state, even with the initial configuration, results in a metastable failure. Therefore, if the triggering event transitions the system into an $NS$ state, initiating a simulation from that state with the original configuration results in a non-stationary simulation, indicative of a metastable failure.

\fontsize{7.8}{7.8}\selectfont
\begin{equation}
    P_{\text{Metastable Failure }} = P ( State \in NS(\mu, \lambda, \mu_0, \tau) | \mu', \lambda', \Delta_{\text{trigger}})
\end{equation}
\normalsize

Here, the probability of a metastable failure is equivalent to the probability of being in any of the non-stationary states, given the conditions of the triggering event.

\newlength{\textfloatsepsave} 
\setlength{\textfloatsepsave}{\textfloatsep}
\setlength{\textfloatsep}{0pt}
\SetAlCapNameFnt{\scriptsize}
\SetAlCapFnt{\scriptsize}
\begin{algorithm}[t]

  \fontsize{7.5}{7.5}\selectfont
  \SetCommentSty{small}
  \caption{Calculate Metastable Failure's Probability}
  \label{alg:alg_ms}
  \DontPrintSemicolon\SetNoFillComment
  \SetKwFunction{procedureName}{\textbf{\ourModelName}} 
  \SetKwData{sset}{Stationary States}
  \SetKwData{pms}{$P_{MS}$}
  \SetKwData{trigger}{Trigger Probabilities}
  \SetKwProg{myalg}{procedure}{}{}    
  \nonl \myalg{\procedureName{$\lambda, \mu, \mu_0, \tau, \lambda', \mu', \Delta_{\text{trigger}}$}}{
  \sset $\gets$ \{\}\;
  $i$ $\gets$ $0$  \hfill {\tcc{Number of requests in orbit space}} \label{alg:alg_ms_p1_s}
  $iBoundry$ $\gets$ $false$ \;
  \While {\textbf{not} $iBoundry$}{
    $j$ $\gets$ $0$ \hfill {\tcc{Number of requests in primary queue}}
    $jBoundry$ $\gets$ $false$ \;
    \While{\textbf{not} $jBoundry$} {
        Simulation $\gets$ $M((i, j), \mu, \lambda, \mu_0, \tau)$ \\
        \If{Simulation \textbf{is not} $\varnothing$}{
            \sset $\gets$ \sset $\cup$ $\{(i, j)\}$
        }
        \Else{
            $jBoundry$ $\gets$ $true$ \;
            \If{j is 0}{
                $iBoundry$ $\gets$ $true$
            }
        }
        $j$ $\gets$ $j+1$ \;
    }
    $i$ $\gets$ $i+1$ \;
  } \label{alg:alg_ms_p1_e}
  \trigger $\gets$ $M((0, 0), \mu', \lambda', \mu_0, \tau, \Delta_{\text{trigger}})$ \; \label{alg:alg_ms_p2_s}
  \If{\trigger is $\varnothing$}{
     \Return{1} \hfill {\tcc{Triggering event put the system into a non-stationary state and MS failure will happen}}
  }
  \Else{
    \pms $\gets$ $0$\;
    \For{State \textbf{in} \trigger \text{with probability} $> 0$ }{
        \If{ State \textbf{not in} \sset}{
            \pms $\gets$ \pms $+$ State.Probability
        }
    }
    \Return{\pms}
  } \label{alg:alg_ms_p2_e}
  }{}

\end{algorithm}
\setlength{\textfloatsep}{\textfloatsepsave}

Algorithm~\ref{alg:alg_ms} represents the algorithmic method of \ourModel\ to determine the probability of a metastable failure.
The first segment (lines~\ref{alg:alg_ms_p1_s}-\ref{alg:alg_ms_p1_e}) identifies all states from which initiating the Monte Carlo simulation with the original configuration does not result in a non-stationary simulation. These states are termed stationary states. 
They are found by iterating over the state space and performing the Monte Carlo simulation for each state. This method identifies the boundary between non-stationary and stationary states and gathers the stationary states for the normal-case setup.

If the triggering event propels the system into a non-stationary state, the system remains in a metastable state even after the triggering event is resolved and the initial configuration is restored. The second segment of the algorithm (lines~\ref{alg:alg_ms_p2_s}-\ref{alg:alg_ms_p2_e}) calculates the probability of transitioning to non-stationary states when a triggering event occurs, which is equivalent to the probability of remaining in a metastable state following the triggering event. This is determined by simulating the triggering event and then adding up the probabilities of transitioning into a non-stationary state.

\cat{Computational Complexity}
The time complexity of algorithm~\ref{alg:alg_ms} is $(O(|\overline{NS(\mu, \lambda, \mu_0, \tau)}| + 1) \times  \Delta(M))$, where $|\overline{NS(\mu, \lambda, \mu_0, \tau)}|$ is the number of stationary states for a system with the initial configuration, and $\Delta(M)$ represents the simulation time.
While the number of stationary states is contingent upon the configuration, it remains finite since only a specific set of states is stationary. 
The duration of the simulation is adjustable and can be determined based on the desired level of accuracy.

\begin{figure*}
    \centering
    \begin{subfigure}[b]{0.33\textwidth}
         \centering
         \includegraphics[width=\textwidth]{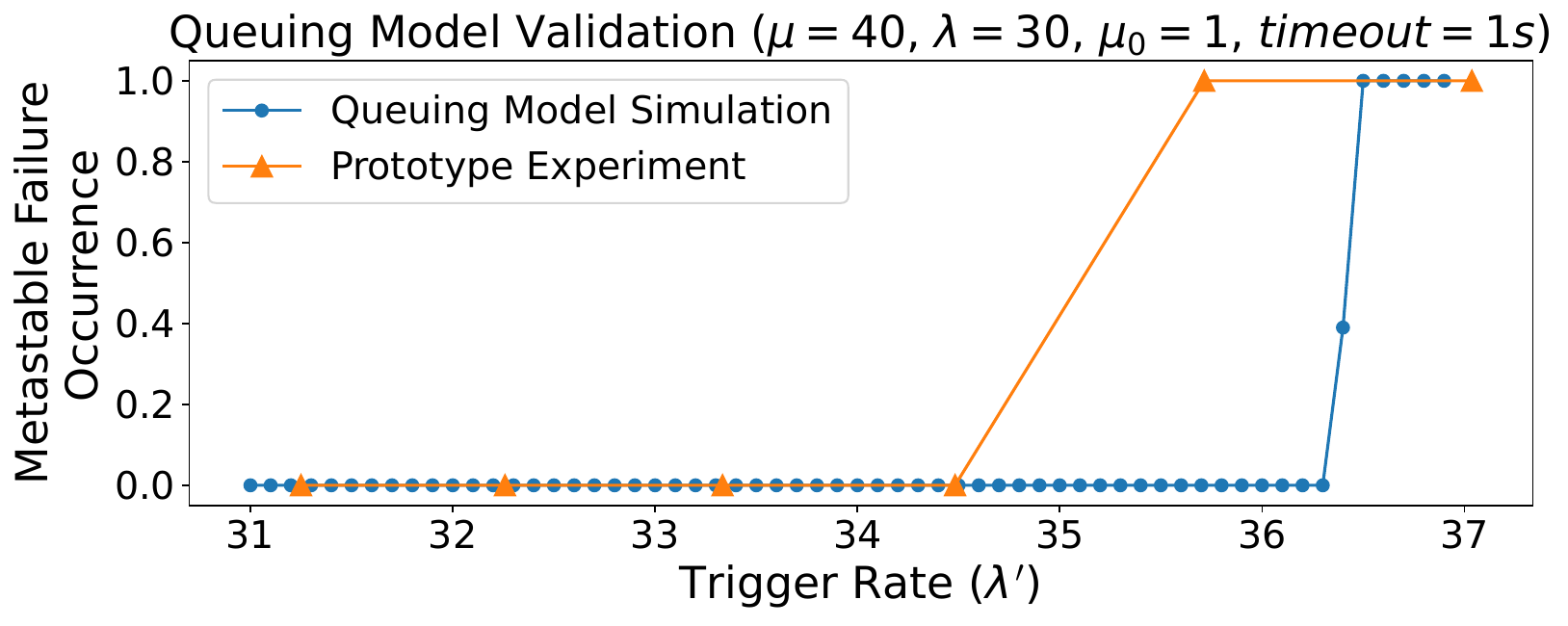}
         \caption{$\text{Batch Size} = 3KB$ }
         \label{fig:validation-3k}
     \end{subfigure}
    \hfill
    \begin{subfigure}[b]{0.32\textwidth}
         \centering
          \includegraphics[width=\textwidth]{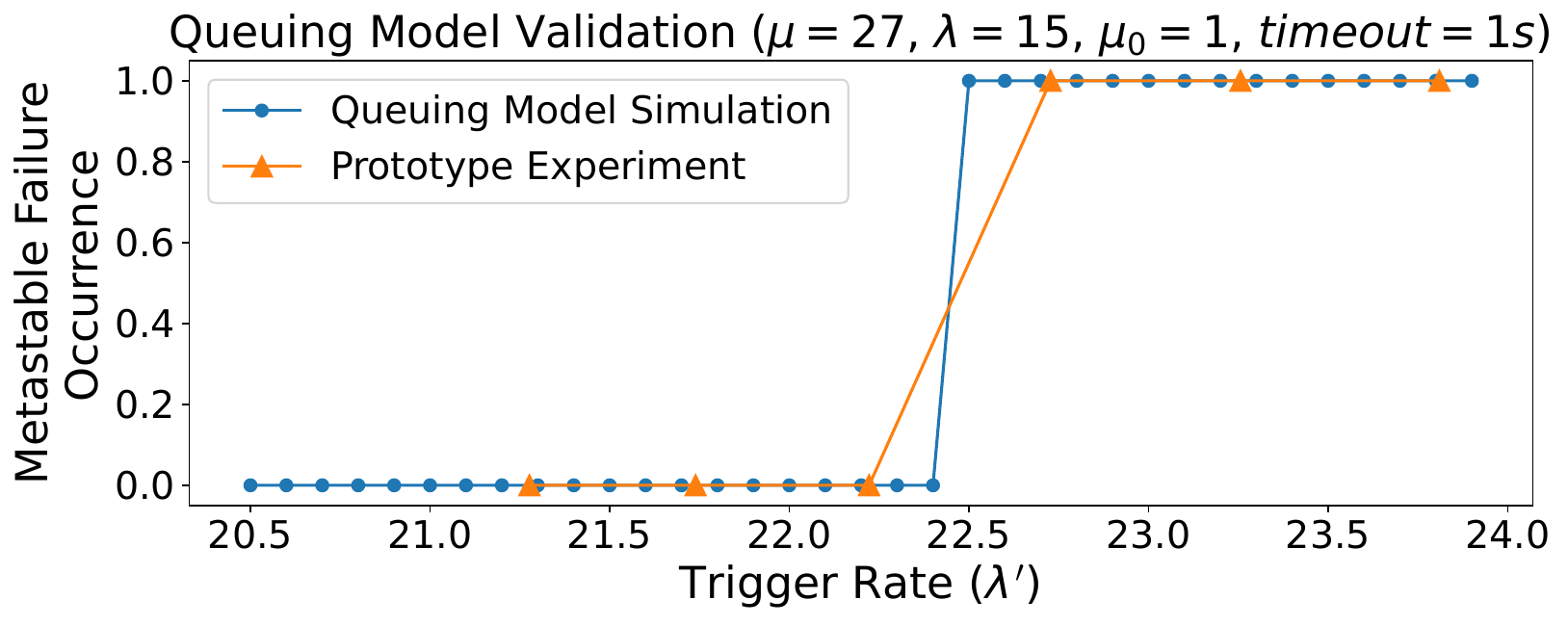}
         \caption{$\text{Batch Size} = 50KB$ }
         \label{fig:validation-50k}
    \end{subfigure}
    \hfill
    \begin{subfigure}[b]{0.33\textwidth}
         \centering
          \includegraphics[width=\textwidth]{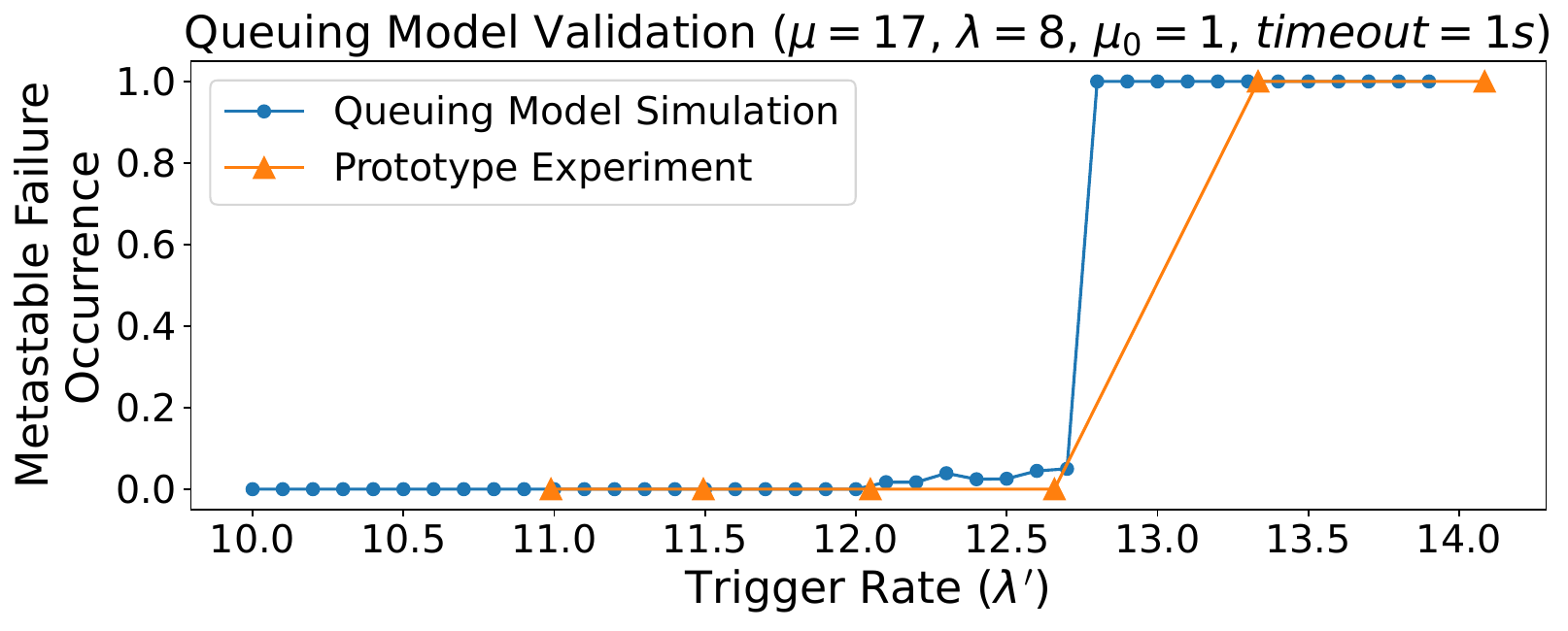}
         \caption{$\text{Batch Size} = 100KB$ }
         \label{fig:validation-100k}
    \end{subfigure}
    \vspace{-0.25em}
    \caption{Queuing model validation with the implemented prototype}
    \label{fig:validation}
\end{figure*}

\subsection{Utility of \ourModelName\ in Real-world} 

Algorithm~\ref{alg:alg_ms}, used as a preprocessing tool in monitoring systems, predicts metastable failures by analyzing network and machine metrics. The \ourModel\ within it assesses the risk of failure based on system capacity and workload changes. If a potential failure is detected, the system adjusts by either suppressing incoming requests or enhancing its capacity to maintain stability. For instance, in a scenario where cache nodes in a system fail, the monitoring system recalculates the capacity-load balance and advises appropriate measures to prevent failure, ensuring operational continuity. A real systems can use \ourModel\ for several beneficial purposes:

\cat{Proactive Failure Prevention} By accurately predicting metastable failures before they occur, the model allows system administrators and engineers to implement preventative measures. Also, the integration of the model with system operations could enable automated responses to emerging failure conditions.

\cat{Enhanced Configuration and Tuning} \ourModel\ allows a system admin to explore whether different configurations and tuning of the system would lead to better resilience to metastable failures. For example, a system admin can test out a range of different configuration metrics to find the best tuning of the system in terms of resilience to metastable failures. 

\cat{Improved System Design} Insights from the model can be used to design more robust and resilient distributed storage systems. A system architect can propose different system features and then modify the \ourModel\ accordingly to match the proposed features. \ourModel\ would provide insights into whether the model would be more or less resilient to metastable failures. For instance, introducing a priority queue to \ourModel\ can model request prioritization and its impact on system resilience to metastable failures.

\section{Validation of \ourModelName}
\label{sec:validation}


To validate our proposed queuing framework, we measure the metastable failure probabilities for various load surge scenarios and compare them with the results from the prototype described in Section~\ref{sec:prototype}. As illustrated in Figure~\ref{fig:validation}, these probabilities were calculated for different load surge scenarios, with each triggering event defined as an increase in incoming load ($\lambda$).

Our prototype experiments are based on the system model described in Section~\ref{sec:system_model} which consists of a replicated storage (PostgresSQL) across three different physical nodes. Consider the first scenario where the batch size is $3KB$ (Figure~\ref{fig:validation-3k}). In this scenario, we assume an average processing rate of 40 requests per second for the prototype, with a maximum batch size of $3KB$. This represents the highest number of $3KB$-size requests that the prototype can handle. Each experiment simulates a load surge scenario, with $3KB$ batch-sized requests sent to the storage system following an exponential distribution to emulate customer arrival patterns. Each request has a timeout of $1$ second and is immediately retried upon timeout, with a maximum retry limit of three times.

In both simulations and prototype experiments, the system's initial incoming load in a stable state ($\lambda$) is set to $30$ for three minutes. Then a triggering event increases the incoming load to a value of $\lambda'$ for one minute ($\lambda'$ is varied in the x-axis.) Following the event, the incoming load returns to its normal state, which is $30$ for another three minutes (Based on the terminology in Section~\ref{sec:scaling-parameter}, $\lambda$ and $\lambda'$ correspond to $sp_1$ and $sp_2$, respectively.)

Figure~\ref{fig:validation-3k} shows the probability of metastable failure for these scenarios using Algorithm~\ref{alg:alg_ms}, compared to the results from the real prototype experiments across three physical machines. According to \ourModel\ results, any load surge triggering event with a rate exceeding $35.5$ has a probability of causing a metastable failure. On the other hand, prototype experiments demonstrate that a triggering event with an equal or higher rate of $35.7$ can lead to a metastable failure, aligning with the probabilities predicted by \ourModel.

Figures~\ref{fig:validation-50k} and~\ref{fig:validation-100k} present similar experiments for batch sizes of $50KB$ and $100KB$, respectively. These experiments differ in the prototype's processing rate as it handles requests of different sizes at different rates. Despite this, the figures show similar results, showcasing a similar occurrence of metastable failures between the prototype and queuing model simulations.

\section{Related Works}
\label{sec:related_works}

\cat{Metastable Failures}
Many studies have investigated the causes of distributed systems failures~\cite{arpaci2001failslow1, gunawi2016whyCloudStop, yuan2014configFailure2, oppenheimer2003confiFailure1, cipar2013straggler1, dean2008straggler2}. A recent study highlighted the possibility of failures arising from the interaction between different (sub-)systems, termed Cross-System Interaction (CSI) failure~\cite{tang2023csifailure}. The authors argue that the reliability of distributed systems is influenced not just by the reliability of individual systems but also by their interconnections. If a metastable failure occurs, its sustaining effect can spread, potentially leading to a CSI failure.

The concept of metastable failure in distributed systems was recently introduced~\cite{bronson2021metastable}. Huang et. al.~\cite{huang2022metastable} expand on the concept of metastable failures by suggesting that systems may become vulnerable not just due to increased load but also due to capacity degradation, introducing varying degrees of vulnerability. Previous works have analyzed metastable failures, but they have not provided a formal methodology for modeling and early detection of them.

Since metastable failures are a recent concept, specific studies on this type of failure are limited. However, various research has investigated different failure types that could trigger a metastable failure or have a relation to it.
Distributed consensus algorithms have addressed crash-failures and aim to enhance system reliability in response to such failures~\cite{du2009multi, ongaro2014raft}. However, a crash-failure can potentially trigger a metastable failure even after the recovery from the crash-failure.
Past research has shown that metastable failures can be triggered by configuration changes~\cite{oppenheimer2003confiFailure1, yuan2014configFailure2} and fail-slow hardware failures~\cite{ arpaci2001failslow1, gunawi2016whyCloudStop}.

\cat{Retrial Queues}
Extensive research has been conducted on retrial queues over the past years~\cite{phung2019retrialsurvey}. In prior studies~\cite{avrachenkov2008retrialqueuenetworks, avrachenkov2010tandemretrial}, a retrial queuing network with a constant retrial rate was employed to model a TCP network. Their model retries a request whenever the server queue is full, leading to blocked customers in the system. Essentially, they consider a network of one or two tandem $M/M/1/\infty$ queues with blocking, and an $M/M/1/\infty$ retrial (orbit) queue.

However, to the best of our knowledge, no work has been done to model scenarios where requests are retried after a specific timeout, regardless of whether they are blocked or not. \ourModel\ captures the timeout retrial nature, which is not modeled in previous works.

In~\cite{neuts1990numericalRetrialQueue}, a numerical approach for analyzing retrial queue models is provided. This method has become the standard for most retrial queuing models to mathematically calculate stationary conditions and their probabilities. The use of numerical analysis may be impractical due to its inherent complexity and the necessary simplifications required in model development. Many works in queuing theory utilized the Monte Carlo method for model analysis~\cite{ouazine2016montecarlo1, dyshlyuk2013montecarlo2}.






\section{Conclusion}

\label{sec:conclusion}

Our work consists of three main contributions to study and analyze metastable failures in replicated storage systems. 
Firstly, we examined various scenarios in which a metastable failure could occur in a replicated storage system driven by a consensus algorithm. Secondly, we reproduced a series of case studies of metastable failures within such a system for further analysis. 
Finally, we proposed, \ourModel, a queuing theory model of metastable failures that allows us to understand whether a system configuration would lead to metastable failures. Experimental validation shows that the theoretical model closely matches the real behavior of metastable failures.
%
Looking forward, \ourModel\ can be generalized to accommodate more complex scenarios, such as systems comprising multiple storage systems. Another future work is the development of mitigation strategies that adjust the rate of requests accepted by the system. Moreover, creating prevention strategies, such as auto-scaling based on the early detection of metastable failures, is a promising next step that can utilize the analytical models we propose.


\section{Acknowledgement}
This research is partly supported by a gift from Roblox and the NSF under grants CNS1815212 and SaTC-2245372.

\bibliographystyle{IEEEtran}

\bibliography{main}


\end{document}